\documentclass{aa}  
\usepackage{graphicx}
\usepackage{txfonts}
\begin{document}

\authorrunning{A. Bonafede et al.}  \titlerunning{The complex radio
  emission from MACS\,J0717+3745} \title{Revealing the magnetic field
  in a distant galaxy cluster: discovery of the complex radio emission
  from MACS\,J0717.5 +3745} \subtitle{}

  \author{A. Bonafede
          \inst{1,2}
	  \and L. Feretti\inst{2}
          \and G. Giovannini\inst{1,2}
	  \and F. Govoni\inst{3}
	   \and M. Murgia\inst{2,3}
           \and G. B. Taylor\inst{4}
           \and H. Ebeling\inst{5}
           \and S. Allen\inst{6}
           \and G. Gentile\inst{7,8}
            \and Y. Pihlstr\"om\inst{4}
          }

\offprints{bonafede@ira.inaf.it} \institute{Universit\`a di Bologna,
  Dip. di Astronomia, via Ranzani 1, I-40126 Bologna, Italy\\ \and
  INAF, Istituto di Radioastronomia, via Gobetti 101, I-40129 Bologna,
  Italy\\ \and INAF, Osservatorio Astronomico di Cagliari, Strada 54,
  Loc.\ Poggio dei Pini, I-09012 Capoterra (Ca), Italy\\ \and
  Department of Physics and Astronomy, University of New Mexico,
  Albuquerque, NM 87131, USA \& Adjunct Astronomer at the National
  Radio Astronomy Observatory\\ \and Institute for Astronomy,
  University of Hawaii, 2680 Woodlawn Drive, Honolulu, HI 96822,
  USA\\ \and Kavli Institute for Particle Astrophysics and Cosmology,
  Stanford University, 382 Via Pueblo Mall, Stanford, CA 94305-4060,
  USA\\ \and Institut d'Astronomie et d'Astrophysique, Facult\'e des
  Sciences, Universit\'e Libre de Bruxelles, CP 226, Bvd du Triomphe,
  B-1050, Bruxelles, Belgium\\ \and Sterrenkundig Observatorium,
  Universiteit Gent, Krijgslaan 281, B-9000 Gent, Belgium.\\ }

\date{Accepted 2 July 2009 Received 18 May 2009}
\abstract
{}
{To study at multiple frequencies the radio emission arising from the
  massive galaxy cluster MACS\,J0717.5+3745 (z=0.55). Known to be an
  extremely complex cluster merger, the system is uniquely suited for
  an investigation of the phenomena at work in the intra-cluster
  medium (ICM) during cluster collisions.}
{We use multi-frequency and multi-resolution data obtained with the
  Very Large Array radio telescope, and X-ray features revealed by
  Chandra, to probe the non-thermal and thermal components of the ICM,
  their relations and interactions. }
{The cluster shows highly complex radio emission. A bright, giant
  radio halo is detected at frequencies as high as 4.8
  GHz. MACS\,J0717.5+3745 is the most distant cluster currently known
  to host a radio halo. This radio halo is also the most powerful ever
  observed, and the second case for which polarized radio emission has
  been detected, indicating that the magnetic field is ordered on
  large scales. }
{}
 \maketitle
\section{Introduction}
A detailed description of the physical conditions and energetics of
the intra-cluster medium (ICM) in galaxy clusters requires adequate
knowledge of the role of the non-thermal components: relativistic
particles and magnetic fields. In recent years, non-thermal ICM
emission and its connection with highly energetic events occurring
during cluster mergers has become a much researched and discussed
topic. Radio halos and radio relics are suspected to be the signature
of mergers, although their origin and evolution is still the subject
of considerable debate. Radio halos have now been observed in the
central region of several clusters of galaxies. With the exception of
the radio halo in Abell 2255 (Govoni et al.\ 2005), they are not
polarized, and their spectrum\footnote{We define the radio spectrum as
  $S(\nu)\propto\nu^{-\alpha}$, where $\alpha$ is the spectral index
  and S the observed flux density at frequency $\nu$} is steep
($\alpha >$ 1).
Radio halo emission is likely due to the re-acceleration of
electrons permeating the cluster volume (see e.g.\ Brunetti et al.\
2008). By contrast, radio relics are usually observed at the periphery
of galaxy clusters.  They vary widely in morphology and size and are
polarized at a level of 20-30\% at 1.4 GHz. They also exhibit steep
radio spectra and are thought to be generated by shocks occurring in
the ICM during merger events (see e.g.\ Roettiger et al.\ 1999; Hoeft
\& Brueggen 2007). Both radio-halo and radio-relic emission indicate
the presence of a $\sim\mu$G magnetic field in the ICM.

In recent years, the presence of magnetic fields in galaxy clusters
has been unambiguously established, and their importance for our
understanding of the physical processes at work in the ICM has been
recognized (see e.g.\ the reviews by Carilli \& Taylor 2002; Govoni \&
Feretti 2004; Ferrari et al.\ 2008; Dolag et al.\ 2008). Magnetic
fields are able to inhibit transport processes like heat conduction,
spatial mixing of gas, and the propagation of cosmic rays. Values of
the Faraday Rotation Measure (RM) have been determined both for radio
galaxies within and behind the cluster, using statistical samples
(e.g.\ Kim et al.\ 1991; Clarke 2004, Johnston-Hollitt et al.\ 2004)
as well as individual clusters by analyzing high-resolution images
(e. g. Taylor \& Perley 1993, Feretti et al.\ 1995, Govoni et al.\
2006). The results are consistent with the presence of magnetic fields
of a few $\mu$G throughout the cluster, in agreement with predictions
from cosmological MHD simulations (Dolag \& Stasyszyn 2008, Donnert et
al.\ 2009). These studies also indicate that the magnetic fields in
the ICM are structured on scales as low as 10 kpc, and possibly even
less. Work on magnetic fields in clusters has, so far, focused on
relatively nearby systems (z$<$0.4), with much less information being
available for clusters at higher redshift. Extending the redshift
range of such studies is crucial because of their importance to the
subject of the formation and evolution cosmic magnetic fields, and to
investigate how the strength and structure of magnetic
fields in clusters is connected to the systems' dynamic history.

MACS\,J0717.5+3745 (MACS\,J0717; $z{=}0.55$) was discovered in the
Massive Cluster Survey (MACS; Ebeling, Edge \& Henry 2001) and has
since been found to be one of the most disturbed galaxy clusters known
at any redshift. It is part of the statistically complete sample of
the twelve most distant MACS clusters, all of which lie at $z{>}0.5$
(Ebeling et al.\ 2007).  In-depth optical and X-ray studies of the
galaxy population and the ICM of MACS\,J0717 identified it as one of
the most promising targets for studies of the physical mechanisms
governing merger events (Ebeling et al.\ 2004; Ma et al.\ 2008, 2009).

Our detailed study of the radio emission arising from both galaxies
and the ICM of this cluster is prompted by the work of Edge et
al.\ (2003) who discovered a radio relic in MACS\,J0717 when analyzing
data from the Faint Images of the Radio Sky at Twenty-cm (FIRST,
Becker et al. 1995 ) survey. Since radio emission is indicative of the
presence of a magnetic field, such observations are the most direct
way to obtain information of this fundamental ingredient in the
physics of the ICM.

In Secs.~\ref{sec:ottico} and \ref{sec:x} we summarize our present
knowledge of this cluster based on previous optical and X-ray
studies. Radio observations and data reduction techniques are
described in Sec.~\ref{par:obs}.  In Secs.~\ref{sec:radio} and
\ref{sec:halo} we discuss the total-intensity emission from the radio
galaxies and from the ICM. In Sec.~\ref{Par:relic} we analyze the
polarization properties of the radio emission, while in
Sec.~\ref{sec:RM} results for the Faraday Rotation are presented and
discussed. The spectral index and the magnetic field properties of
this cluster are the subjects of Secs.~\ref{sec:spix} and
\ref{sec:MF}. Finally, conclusions are presented in
Sec.~\ref{Sec:res}.  We use the concordance cosmological model
$\Lambda$CDM, with H$_0=$71 km/s/Mpc, $\Omega_M=$0.27, and
$\Omega_{\Lambda}=$0.73. In this cosmology, at redshift z=0.55, 1
arcsecond corresponds to a scale of 6.394 kpc.

\begin{figure*}
\includegraphics[width=0.95\textwidth]{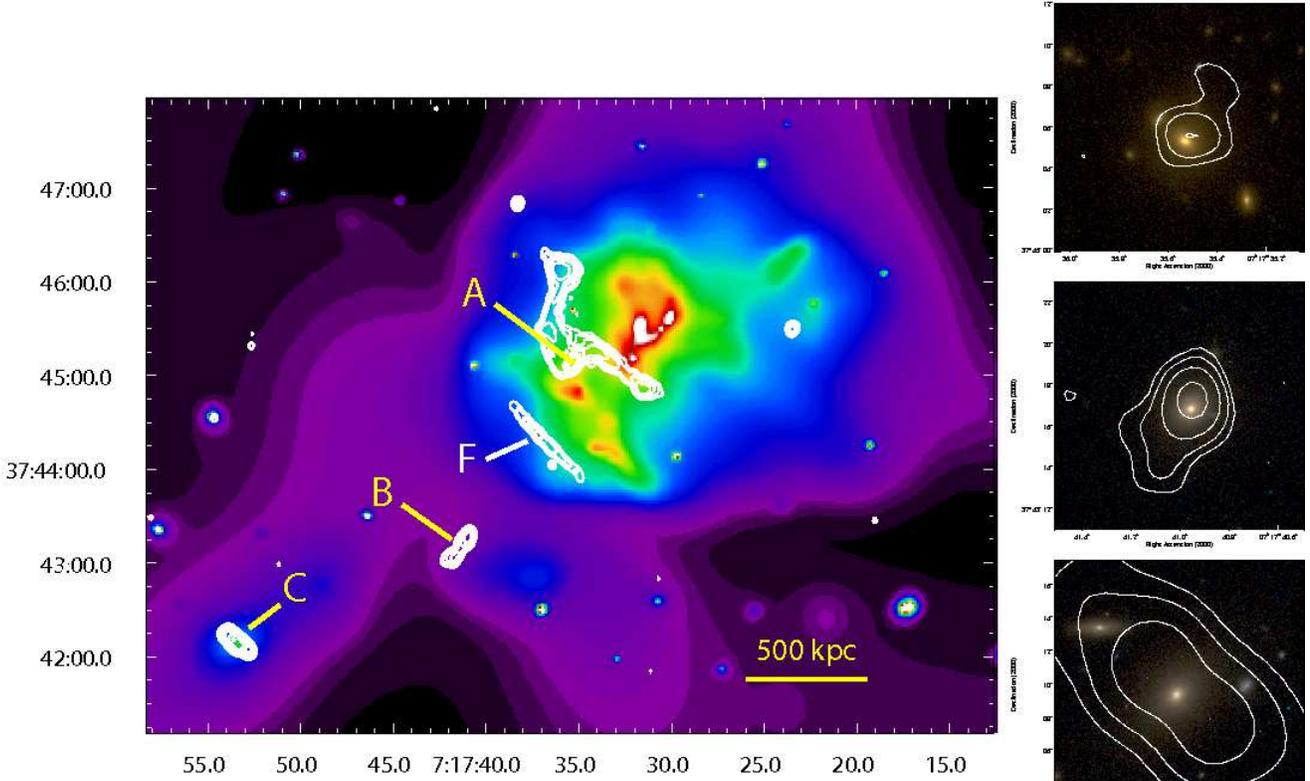}
\vspace{5pt}
\caption{Colors represent the X-ray surface brightness of MACS\,J0717
  as observed with Chandra in the 0.5--7 keV band, adaptively smoothed
  using the {\it asmooth} algorithm (Ebeling et al.\ 2006) requiring a
  minimal significance of 99\% with respect to the local
  background. Contours show the emission from radio sources at 1.365
  GHz.  The size of the restoring beam is 5$''\times$4$''$, and the rms noise is
  $\sigma\sim$ 3.0 $\mu$Jy/beam. The lowest contour level corresponds
  to 5$\sigma$, additional contours are spaced by factors of 2. The
  source labelled F is a foreground radio galaxy. The side panels show
  contours of the radio emission from sources A, B, and C (top to
  bottom), overlaid on optical images obtained with the Hubble Space
  Telescope (HST). For sources A and B, we show the emission at 8.460
  GHz, where the restoring beam is 2.4$''\times$2$''$, and the rms
  noise $\sigma\sim$0.04 mJy/beam. The first contour is placed at the
  3$\sigma$ level; additional contours are spaced by factors of 2.
  Radio emission from source C is shown at 1.365 GHz, using the same
  contour levels as in the large image. The HST images were obtained
  with the Advanced Camera for Surveys (ACS), using the F555W and
  F814W filters (source A), and the F606W and F814W passbands (source
  B and C).}
\label{fig:radioX}
\end{figure*}

\begin{table*}
\label{tab:radioobs}
  \caption{Radio observations}
 \centering
  \begin{tabular}{@{}lccccccccl@{}}
  \hline
Project&   RA   & DEC     & $\nu$ & Bandwidth& Configuration&Date& ToS& Restoring beam &rms noise \\
Id &(J2000) & (J2000) & (GHz) & (MHz)    &         &        & (Hours)  &  $''\times''$    & (mJy/beam)\\
 \hline
AG761& 07h17m35.0s&37d45$'$07$''$&1.365 - 1.435 & 25&  B      &16-DEC-07&2.2&5$\times$4&0.030\\
AG761& 07h17m35.0s&37d45$'$07$''$&1.485 - 1.665 & 25&  B      &24-DEC-07&2.8&5$\times$4&0.025\\
AG761& 07h17m35.0s&37d45$'$07$''$&4.885 - 4.535 & 50&  C      &24-MAR-08&2.4&5$\times$4&0.015\\
AG761& 07h17m53.0s&37d42$'$11$''$&4.885 - 4.535 & 50&  C      &20-APR-08&2.2&5$\times$4&0.016\\
AT0358& 07h17m30.9s&37d45$'$30$''$&1.425 & 50   &  C  &15-MAR-08&1.8&18$\times$14&0.020 \\
AH748 &07h17m33.8s&37d45$'$20$''$&4.860 & 50&  D      &27-NOV-01&1.5&18$\times$14&0.020\\
AE0125&07h17m35.4s&37d45$'$07$'$&8.460  & 50 &   C    &22-NOV-1998&0.2&2.4$\times$2&0.04\\
\hline
\end{tabular}
\end{table*}

\section{Optical observations}
\label{sec:ottico} 

The complex optical morphology of MACS\,J0717 was first noted by Edge
et al.\ (2003), based on imaging in the V, R, and I passbands obtained
with the University of Hawaii 2.2m telescope. Much deeper observations
conducted since with SuprimeCam, the wide-field imager at the prime
focus of the Subaru 8m telescope, firmly established MACS\,J0717 as a
highly disturbed merger and led to the discovery of a 6-Mpc long
filament leading into the cluster from the South-East (Ebeling et al.\
2004). Spectroscopic observations of over a thousand galaxies in the
field of MACS\,J0717 have been performed in order to probe its spatial
and kinematic structure along the line of sight, and to characterize
the galaxy population as a function of cluster environment (Ma et al.\
2008). Finally, space-based observations with the Advanced Camera for
Surveys (ACS) aboard the Hubble Space Telescope (GO-09722, PI Ebeling)
provided a high-resolution view of MACS\,J0717, including the
interface region where the filament meets the dynamically most active
central region of the cluster.

\section{X-ray observations}
\label{sec:x}

With an X-ray luminosity of (2.74$\pm$ 0.03 )$\times$10$^{45}$ erg/s
in the 0.1-2.4 keV energy band (Ebeling et al.\ 2007) MACS\,J0717 is
one of the most X-ray luminous clusters known at z$>$0.5.  The cluster
was observed with the ACIS-I instrument aboard the Chandra X-ray
Observatory for a total exposure time of 60 ks (ObsID 4200). A
detailed study of the system's X-ray properties was recently performed
by Ma et al.\ (2008, 2009). Their spatial description of the gas
distribution uses a $\beta$-model (Cavaliere \& Fusco-Femiano 1976):

\begin{equation}
\rho_{gas}=\rho_0 \left[1+\frac{r^2}{r_c^2}\right]^{\frac{-3\beta}{2}}
\end{equation}

where $\rho_{gas}$ is the gas density, $r$ is the radial distance from
the cluster center, and $r_C$ is the cluster core radius. By fitting
this model to the X-ray surface brightness data, they derived
$\beta=1.1\pm 0.1$, $r_c=92''\pm 6''$ and $\rho_0=1.71\pm 0.05\times
10^{14}M_{\odot}Mpc^{-3}$.  Ma and co-workers report the detection of
X-ray emission from the filament and, for the main cluster, a very
complex X-ray morphology, including dramatic variations in the
intra-cluster gas temperature, with extreme values of 5 and over 20 keV
(the average gas temperature quoted by Ebeling et al.\ (2007) is 11
keV). The authors' joint optical/X-ray analysis of all available data
identifies the filament as the source of both continuous and discrete
accretion of matter by the cluster from a south-easterly direction,
and isolates four distinct subclusters participating in an ongoing
triple merger. The X-ray emission from MACS\,J0717 and the
cluster-filament interface is shown in Fig. \ref{fig:radioX}.

\section{Radio observations and data reduction}
\label{par:obs}

We investigate the radio emission from MACS\,J0717 with
multi-frequency and multi-resolution VLA observations. Specifically,
we performed new high-resolution observations in full-polarization
mode. These observations focused on radio galaxies in the field, as
well as on the relic, to study their Faraday Rotation Measure. In
addition, new low-resolution observations were performed to study the
diffuse emission of the cluster. We also used archival VLA
observations, both of high and low resolution, as specified below.
\subsection{High-resolution observations} 
MACS\,J0717 was observed with the B array at four frequencies within
the 20-cm band (1.365 GHz, 1.435 GHz, 1.485 GHz, and 1.665 GHz),
and with the C array at two frequencies within the 6-cm band (4.535
GHz, and 4.885 GHz). The targets of these observations were the relic,
the radio galaxy embedded in the relic emission (labelled A in
Fig.~\ref{fig:radioX}) and the radio galaxies labelled B and C in
Fig.~\ref{fig:radioX} detected at larger projected distance from the
cluster center.  Two separate pointings were necessary at 6 cm to
avoid bandwidth and primary-beam attenuation. The source 0137+331
(3C48) was used as the primary flux-density calibrator, and the source
0521+166 (3C138) as an absolute reference for the electric vector
polarization angle. The nearby source 0713+438 was observed at
intervals of $\sim$20 min and used as phase calibrator. Calibration
and imaging were performed with the NRAO Astronomical Image Processing
System (AIPS), following standard procedures. Self-calibration was
performed to refine antenna phase solutions, followed by a final gain
and amplitude self-calibration cycle. Images of the total intensity
(Stokes I), as well as of the Stokes parameters U and Q, were produced
for each frequency separately. We then derived images of the polarized
intensity $P=\sqrt{(Q^2+U^2)}$ and of the polarization angle $\Psi=0.5
\arctan(U/Q)$. These images were restored with a Gaussian beam of
FWHM$=$5$''\times$4$''$ which corresponds to a linear resolution of
$\sim$32 kpc.  In order to distinguish the relic emission from that of
embedded radio sources we have retrieved from the NRAO archive a short
($\sim $ 10 min) observation performed at 8.460 GHz (project ID
AE125). Here the source 3C147 was used as primary flux-density
calibrator, and the source 0741+312 was used as phase
calibrator. Because of the smaller field of view, only the sources A
and B are visible in the radio image. Observational details are
reported in Table \ref{tab:radioobs}.

\begin{figure*}
\includegraphics[width=0.95\textwidth]{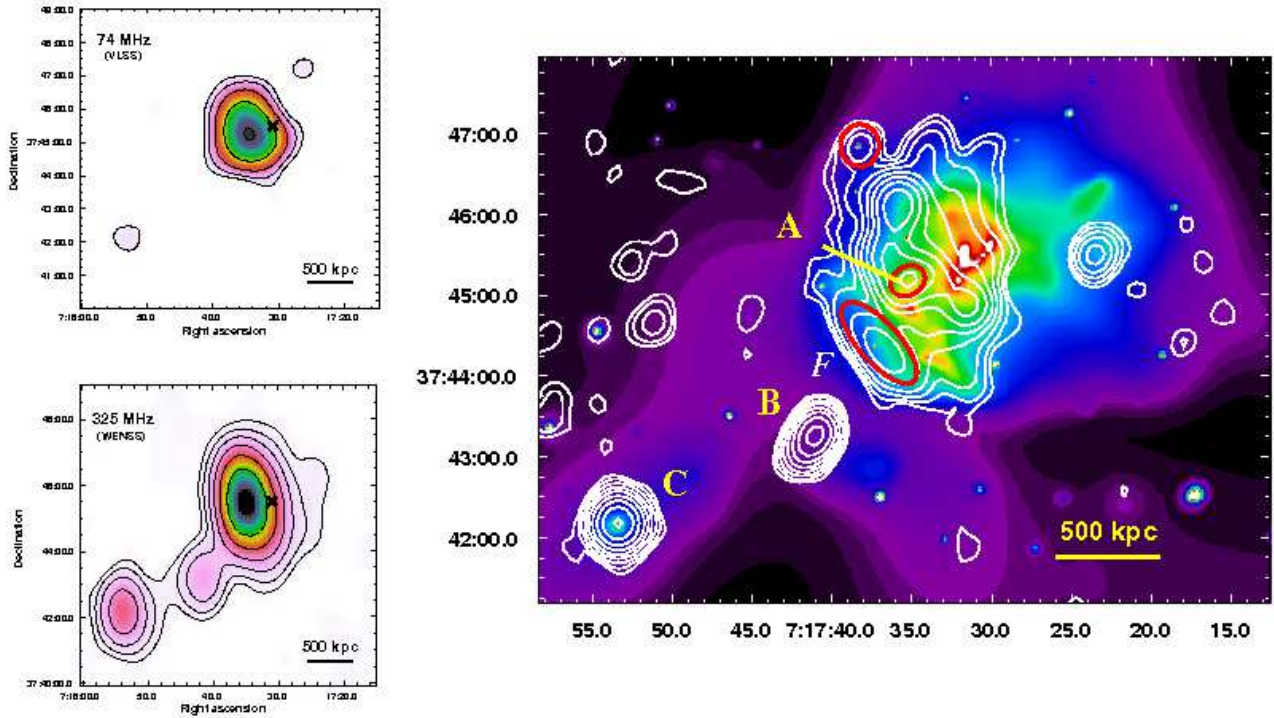}
\caption{Radio emission from MACS\,J0717 at 74 MHz from the VLSS (top
  left), and at 325 MHz from the WENSS (bottom left). Contour levels start
  at 3$\sigma$ (i.e.\ 0.24 Jy/beam for the VLSS image and 9 mJy/beam for
  the WENSS image) and are then spaced by factors of 2. The cross
  marks the X-ray center. See text for details. Right: Contours of
  the radio emission at 1.425 GHz and at low resolution, overlaid on a
  color image of the X-ray emission as observed with Chandra (see
  caption of Fig.~1). Contour levels start at 9 mJy/beam (3$\sigma$);
  subsequent contour levels are spaced by factors of 2. The FWHM of the 
  restoring beam is $\sim$21$''\times$18$''$. Circles and ellipse mark 
  sources embedded in the radio-halo emission.}
 \label{fig:comboHalo}
\end{figure*}

\subsection{ Low-resolution observations} 

MACS\,J0717 was observed with the C array at 1.425 GHz in order to
properly image the extended emission. The source 1331+305 (3C286) was
used as primary flux-density calibrator, and also as an absolute reference
for the electric vector polarization angle. The source 0713+438 was
used as phase calibrator.  Calibration and imaging were performed with
the NRAO Astronomical Image Processing System (AIPS), following
standard procedures. In addition, we recovered from the VLA data
archive an observation at 4.860 GHz (project ID AH748). Here the source
0714+335 was used as phase calibrator. At  
both frequencies total-intensity maps (Stokes I) were produced using  
natural weighting and restored with the same beam of FWHM
$\sim$18$''\times$14$''$, corresponding to $\sim$100 kpc. In order to
study the diffuse polarized emission, I, Q, and U images were also
produced from 1.425 GHz observations at low resolution (FWHM
$\sim$21$\sim$18$''$).  Observational details are reported in
Table \ref{tab:radioobs}.

\begin{table*}
  \caption{Total and polarization-intensity radio emission. }
   \centering
\begin{tabular}{lccccccc}
  \hline
Name& RA & DEC & dist &$\nu$&Peak brightness  &Flux density & Peak of pol. flux\\
    & (J2000) & (J2000) & kpc& GHz& (mJy/beam)       & (mJy)  & (mJy/beam)\\ 
 \hline
Source A&07h17m35.4s& +37d45'08$''$&450  &1.365&8.3&22.2&0.3\\
        &&&&1.435&8.4&22.4&0.3\\
        &&&&1.485&7.7&20.0&0.2\\
        &&&&1.665&7.4&18.1&0.3\\
        &&&&4.535&2.3&4.7&0.2\\
        &&&&4.885&2.1&3.9&0.2\\
        &&&&8.460&0.5&0.7&\\
Source B&07h17m40.9s& +37d43'17$''$&1150&1.365&9.2&19.9&0.2\\
        &&&&1.435&9.3&18.8&0.3\\
        &&&&1.485&8.8&18.3&0.4\\
        &&&&1.665&8.8&18.3&0.4\\
        &&&&4.535&4.1&7.0&0.3\\
        &&&&4.885&3.8&6.4&0.2\\
        &&&&8.460&1.6&1.9&\\
Source C&07h17m53.7s& +37d42'11$''$&2140&1.365&10.0&32.8&0.1\\
        &&&&1.435&9.9&32.3&0.2\\
        &&&&1.485&9.6&30.5&0.2\\
        &&&&1.665&9.8&30.4&0.2\\
        &&&&4.535&5.4&17.2&0.5\\
        &&&&4.885&5.0&14.7&0.4\\
\hline
\end{tabular}
\label{tab:sourcesPol}
\end{table*}

\section{Radio emission: the relic-filament structure and the radio galaxies}
\label{sec:radio}

The presence of non-thermal emission associated with the ICM in
MACS\,J0717 was first reported by Edge et al.\ (2003) who classified
the emission as a relic.  The wealth of radio data described in the
previous section allows a multi-resolution and multi-frequency study
of the radio emission arising from this cluster.
Fig.~\ref{fig:radioX} shows the 1.365 GHz data from VLA B-array
observations in contours, overlaid on the X-ray surface brightness in
colors. The radio data reveal a bright, filamentary structure whose
central part is coincident with the relic discovered by Edge et
al.\ (2003). In addition, several radio sources are detected. Those
related to the cluster and the large-scale optical and X-ray filament
are labelled A, B, and C, whereas the source labelled F is a radio
galaxy in the foreground ($z=0.1546$, Ma et al., in preparation).

A proper study of the extended emission associated with the ICM needs
to take into account possible contamination from radio galaxies
embedded in the diffuse emission. This can be done with the help of
high-resolution and high-frequency observations in which the extended
emission is resolved out and too faint to be detected. Using 8.460 GHz
observations we have identified one such embedded radio galaxy,
labelled A in Fig.~\ref{fig:radioX} and shown in a close-up view in
the side panel of the same figure. This source is $\sim$70$''$ away
from the cluster center in projection. It is consists of a bright
region, likely the core of the radio galaxy, with a spectral index of
$\sim$0.7 between 1.465 GHz and 4.885 GHz, and an extended, more
diffuse region extending toward the NW. The spectral index steepens up
to $\sim 2$ with increasing distance from the core. The optical
counterpart coincides with the radio core (Fig.~\ref{fig:radioX}) .

Moving out from the cluster center, two additional radio galaxies,
labeled B and C in Fig.~\ref{fig:radioX}, are visible at all the observed
frequencies. Both are located to the SE of the cluster's X-ray center,
at projected distances of $\sim$ 180$''$ and $\sim$ 335$''$,
respectively. We note their position along the X-ray and optical
large-scale filament detected by Ebeling et al.\ (2004). Source C is
the brightest cluster galaxy of the next cluster (also detected in the
Chandra observation) that is going to merge with MACS\,J0717. In
Table~\ref{tab:sourcesPol} we report the main radio properties of the
identified radio galaxies at the observed frequencies. The optical
counterparts of A, B, and C are all spectroscopically confirmed to lie
at redshifts consistent with that of MACS\,J0717 proper.

\section{The radio halo}
\label{sec:halo}

Being sensitive to structure on larger angular scales, observations
performed with the C array at 1.425 GHz and with D array at 4.860 GHz
reveal the presence of an extended radio halo permeating the cluster
volume around the filamentary structure visible at high resolution and
discussed in the previous section. MACS\,J0717 is the most distant
cluster in which a radio halo has been observed so far and its
presence, together with the radio halo detected in the cluster CL0016
by Giovannini \& Feretti (2000) at z=0.54, indicates that the ICM is
already significantly magnetized at redshift z$\sim$0.5.  The
detection of yet more extended emission around the filamentary
structure detected at higher resolution raises some questions about
the nature and origin of the latter. This structure could either be a
radio relic located at the cluster periphery, but appearing close to
the cluster center when viewed in projection, or it could be a
filamentary feature that is in fact part of the radio halo. We will
further investigate these hypotheses with the help of additional
information on the polarization and spectral index of the radio emission
(see Secs.~\ref{sec:RM} and \ref{sec:spix}). In the following
analysis, we will refer to this feature as the relic-filament in order
to stress its uncertain nature.

In Fig.~\ref{fig:comboHalo} the halo emission at 1.425 GHz is shown
overlaid onto the cluster X-ray emission. The maximal angular extent
of the halo at 1.425 GHz is $\sim$240$''$in the NS direction,
corresponding to a linear size of $\sim$1.5 Mpc.
At 4.860 GHz only the brightest regions of the halo are visible, and
its angular extent is reduced to $\sim$160$''$ (i. e. $\sim$ 1 Mpc ).

As shown in the panels on the left of Fig.~\ref{fig:comboHalo}, the
radio halo in MACS\,J0717 is also detected at 74 MHz in the VLSS (VLA
Low Sky Survey, Cohen et al.\ 2007), and at 325 MHz in the Westerbork
Northern Sky Survey (WENSS, Rengelink et al.\ 1997). The VLSS was
performed with the B array at a resolution of 80$''\times$80$''$ and
with an rms noise level of $\sim$0.08 Jy/beam, while the WENSS has a
resolution of 54$''\times$54$'' cosec(DEC)$, which translates into
54$''\times$84$''$ for the declination of MACS\,J0717.  The apparent
angular extent of the halo in the WENSS image is $\sim$280$''$,
corresponding to $\sim$1.8 Mpc. Furthermore, faint radio emission that
appears to connect the radio galaxies B and C with the central halo is
detected at 325 MHz at 3 $\sigma$ significance. We note that this
region coincides with the large-scale filament funneling matter onto
MACS\,J0717 that has been detected at optical and X-ray wavelengths by
Ebeling et al.\ (2004) and Ma et al.\ (2009). The detection of radio
emission in this area might indicate that the magnetic field is
already present in the filament before the amplification due to the
merger process has occurred. However, the feature is detected at
3$\sigma$ significance and could simply be the result of blending of
the two radiosources B and C at the low resolution of the 325MHz
data. Deeper observations would be required to clarify this issue.
\begin{figure*}
 \includegraphics[width=17cm]{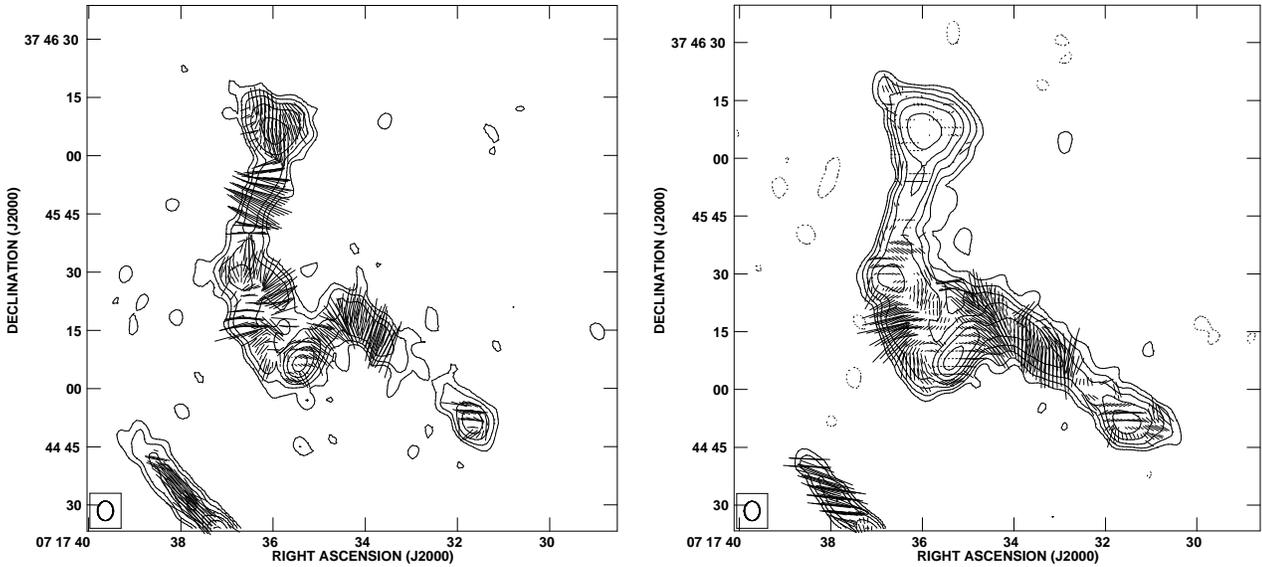}
 \caption{Polarized emission from MACS\, J0717 at 4.885 GHz (left
   panel) and at 1.365 GHz (right panel). Contours represent the total
   intensity. The FWHM of the restoring beam is 5$''\times$4$''$. The
   lowest contours shown are at --3$\sigma$ and 3$\sigma$; subsequent
   contours are spaced by factors of 2.  Lines represent the
   polarization vectors: line orientation indicates the direction of
   the E field, while line length is proportional to the polarization
   percentage. 1$''$ corresponds to 3\%.}
\label{fig:pol}
\end{figure*}

\begin{figure*}
\includegraphics[width=0.95\textwidth]{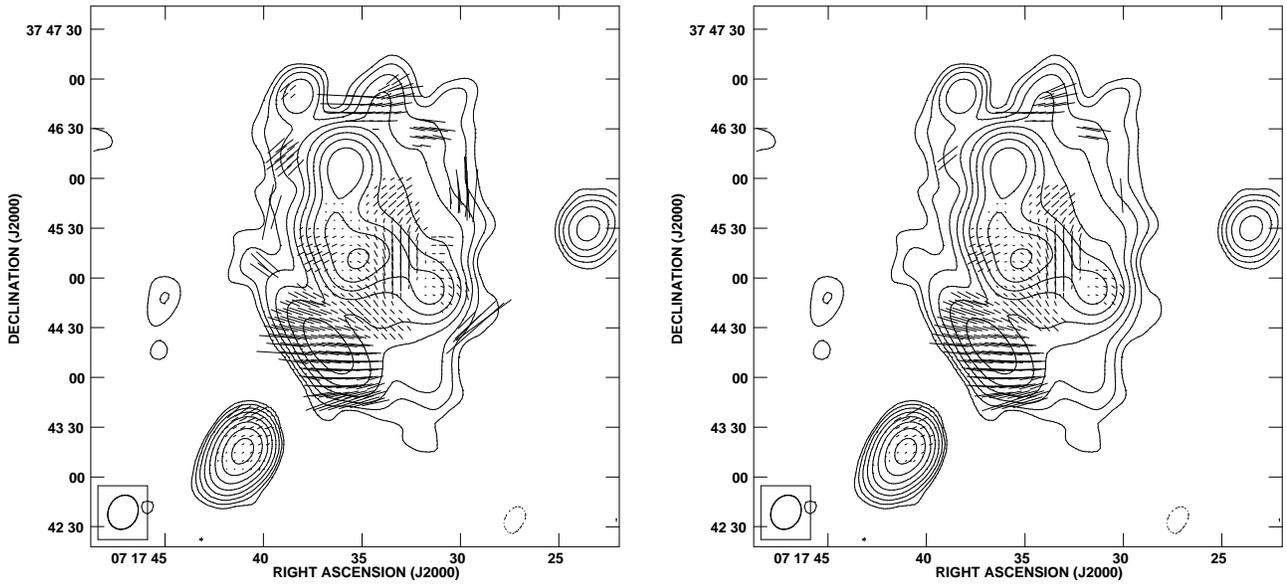}
 \caption{Polarized emission from the cluster at 1.425 GHz. Contours
   show to the total-intensity emission at the resolution of
   21$''\times$18$''$, starting at 3$\sigma$, with higher contour
   levels spaced by factors of 2. Lines refer to the E vectors. Their
   orientation represents the projected E-field not corrected for
   Galactic rotation. Their length is proportional to the fractional
   polarization: 1$''$ corresponds to 13\%.  Fractional
     polarization pixels with a signal-to-noise ratio less than 3
     (left) and 5 (right) were blanked (see text).}
\label{fig:Pol_Halo}
\end{figure*}
\section{Polarized emission from the ICM}
\label{Par:relic}

High-resolution observations were performed in full-polarization mode,
as well as 1.425 GHz observations at low resolution. We are thus able
to study the polarized emission in a wide range of resolutions, and,
at high resolution, in a wide range of frequencies.

\subsection{High-resolution images}

The relic-filament is located $\sim$40$''$ ($\sim$ 260 kpc in
projection) SE of the main X-ray peak of the cluster. Its maximal
angular extent is $\sim$130$''$ at 1.4 GHz, corresponding to $\sim$
830 kpc at the cluster redshift, and its orientation is
$\sim$30$^{\circ}$.  Its flux density (excluding the region covered by
source A) is $\sim$83 mJy at 1.365 GHz, and $\sim$16 mJy at 4.885 GHz.

The relic-filament is polarized at all of the observed frequencies.
Its mean fractional polarization is 8\% at 1.365 GHz and increases to
17\% at 4.885 GHz. These values are consistent with those found in
other relics, as well as with the values found in the filament of the
radio halo in Abell 2255 (Govoni et al.\ 2005). In Fig.~\ref{fig:pol}
the E-vectors at 1.365 and 4.885 GHz are shown.  We note that the
polarization of the relic-filament is not uniform.  Its northern part
is strongly depolarized at 1.365 GHz. Its mean fractional polarization
is a few percent at 1.365 GHz and increases to 20\% at 4.885
GHz. Depolarization between 1.365 and 4.885 GHz also occurs in the
southern part, as expected, but is less dramatic (the mean fractional
polarization is $\sim$9\% at 1.365 GHz and $\sim$16\% at 4.885 GHz).

\subsection{Low-resolution images}

In order to reveal any extended emission from the radio halo, an image
was produced using a Gaussian taper to down weight long-baseline data
points in the UV plane. The image was then restored with a Gaussian
beam of FWHM $\sim$21$''\times$18$''$ (see Fig.~\ref{fig:comboHalo}).
\\ In order to investigate the presence of polarized emission from the
radio halo we produced Stokes Q and U radio images following the same
technique. We then derived the polarization angle image and the
polarization intensity image without imposing any cut. From the
polarization intensity image we derived the fractional polarization
image by dividing the polarization intensity image to the total
intensity image, and we considered as valid pixels those whose
signal-to-noise ratio was $>$3, and $>$5 in the output image. The cut
on the final image, done on the basis of the signal-to-noise ratio, is
done to get rid of possible spurious polarization. The resulting
images are shown in Fig. \ref{fig:Pol_Halo}. From them we can gather
that there is a detection of polarization in the halo, mostly
concentrated in one region in the center, and also strong at the
edges, though this becomes weaker when the cut on the fractional
polarization images are more severe. \\ From Fig.  \ref{fig:Pol_Halo}
we can also gather some indication that the radio emission detected at
high resolution and previously classified as relic is likely a
polarized filament belonging to the radio halo. We note, in fact that
the polarized structure as revealed from the polarization vectors does
not show any jump between the relic and the more extended part, but
instead the E-vectors trace with continuity the brightest part of the
radio halo. We refer, for comparison to the case of Abell 2256 (Clarke
\& Ensslin 2006). Here the polarized emission image marks a clear and
sharp distinction between the radio halo and the radio relic. Thus, we
report the polarization percentage values of the whole ICM emission
(excluding only the contribution of source A).  The mean polarization
percentage at 1.425 GHz is $\sim$2-7\%, (3-5$\sigma$ detection) with
lower value in the central part of the halo, that is $\sim$0.01-0.6\%
(3-5$\sigma$ detection) and higher values at the edges, where it
reaches a maximum value of $\sim$24-34\% (3-5$\sigma$ detection). 
  We note that values reported here based on low-resolution
  observations might be affected by beam depolarization. Indeed,
  small-scale variations of the magnetic-field orientation are evident
  from high-resolution images (see Fig.~\ref{fig:pol}). At 20 cm we
  get 1 radian of rotation for a RM of 25 rad$/m^2$ (see
  Eq. \ref{eq:rm}), thus for a 20$''$ beam the RM gradient is about 1
  rad$/m^2/''$ to cause cancellation within the beam, and any
  reasonable ICM model could produce this. Therefore the mean
  polarization percentages should be considered lower limits.

\begin{figure*}[ht]
 \includegraphics[width=0.95\textwidth]{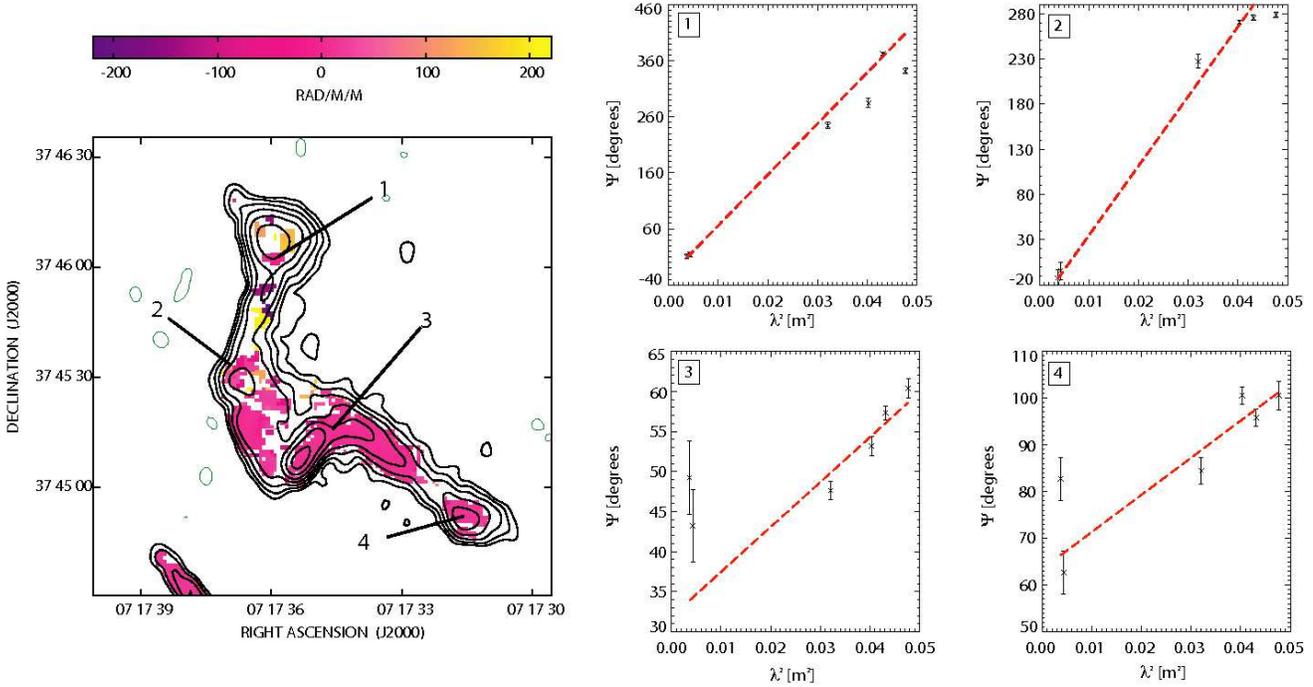}
 \caption{Left panel: Contours refer to the radio emission at 1.425
   GHz. The restoring beam FWHM is 5$''\times$4$''$. Colors represent
   the attempt to fit the Polarization Angle at the observed
   frequencies with the $\lambda^2$ law in the relic-filament
   region. Results from fits of a $\lambda^2$ law to the polarization
   angle are shown in the plots on the right. They refer to four
   random pixels located in different parts of the relic-filament and
   of the source A as indicated in the left panel.}
\label{fig:RMrelic}
\end{figure*}

\section{Rotation Measures}
\label{sec:RM}
Since synchrotron radiation is linearly polarized, its interaction
with the magnetized ICM results in a rotation of the wave polarization
plane, called Faraday Rotation. If the radiation passes through an
external screen, the observed polarization angle $\Psi_{obs}$ at a
wavelength $\lambda$ is related to the intrinsic one $\Psi_{int}$ by
the relation
\begin{equation}
\label{eq:rm}
\Psi_{obs}=\Psi_{int}+\lambda^2\times RM.
\end{equation}
Here RM is the Rotation Measure, which is related to the gas density $n$ and
to the magnetic field intensity along the line of sight according to:
\begin{equation}
\label{eq:rm2}
RM \propto \int_{los}nB_{//}dl.
\end{equation}
The situation in this cluster could however be much more complicated
that the drawn picture.  If radiation is strongly affected by beam
depolarization, or if the Faraday rotation is internal,
Eq. \ref{eq:rm} is not applicable (see Burn 1966).  The large
frequency range of our observations allows us to test whether a simple
linear fit of the polarization angle versus $\lambda^2$ is consistent
with our data.\\

\subsection{Rotation Measure fits}

We performed a fit of the polarization angle images as a function of
$\lambda^{2}$ for the relic-filament, and for the sources B and C. We
used the PACERMAN algorithm (Polarization Angle CorrEcting Rotation
Measure ANalysis) developed by Dolag et al. (2005).  The algorithm
solves the n$\pi$ ambiguity in low signal-to-noise regions exploiting
the information of nearby reference pixels, under the assumption that
the reference pixel is connected to the nearby areas as far as the
polarization angle gradient is under a certain threshold in all of the
observed frequency maps simultaneously. We considered as reference
pixel those which have a polarization angle uncertainty smaller than 7
degrees, and fixed the gradient threshold to 15 degrees. 7 degree
error of the polarization angle corresponds to 3$\sigma$ level in both
U and Q polarization maps simultaneously.\\ We note that some regions
show a high polarized flux at 6cm while they are weakly polarized at
20cm. This could indicate that the Faraday rotation is higher in these
regions, leading to higher depolarization in lower frequency
observations. Excluding these regions would correspond to a bias
toward low RM regions. So we decided to allow PACERMAN to perform the
RM fit if at least in 3 frequency maps the above mentioned conditions
were satisfied.\\ Once the RM image is obtained, the galactic
contribution to the observed RM must be subtracted.  We computed the
average RM for extragalactic sources from the catalog published by
Simard-Normandin et al. (1981). The cluster's galactic coordinates
are: lon$=180.25^{\circ}$ and lat$=+21.05^{\circ}$. It results that in
a region of 15$^{\circ}\times$15$^{\circ}$ centered on the cluster the
Galactic contribution is $\sim17rad/m^2$.  This value is also
consistent with the $\langle RM \rangle$ that we find in our
observations for the foreground galaxy F located at $07^h17^m37.2^s$,
$+37^{\circ}44'21''$ (J2000). Its $\langle RM \rangle$ is
15$\pm3rad/m^2$, with values going from -3$rad/m^2$ to 28$rad/m^2$. \\

\begin{table}
 \caption{RM values for sources B and C }
 \begin{tabular}{lcccc}
\hline
Name&   $\langle RM_{obs} \rangle$ & $\langle RM_{crf} \rangle$     & Fit error (average)& N of beams\\
    &   rad/m$^2$  &    rad/m$^2$ & rad/m$^2$     &\\
\hline
Source B & -130  & -312 &1 & 3\\
Source C &  85   & 204  &1 & 3 \\
\hline
\multicolumn{4}{l}{Col. 1: Source name ; Col 2: observed $\langle RM\rangle$;}\\
 \multicolumn{4}{l}{Col 3: $\langle RM\rangle$ in the cluster rest
   frame; }\\
 \multicolumn{4}{l}{Col. 4: fit error; Col 5: number of sampled beams}\\
\end{tabular}
\label{tab:RMsource} 
\end{table}

\subsubsection{ \bf RM fit in the relic-filament region}

From the existing data we can only derive the position of the
relic-filament in projection, but not where it lies with respect to
MACS\,J0717 along the line of sight. 
It could be a foreground structure (case 1), a background
structure (case 2) seen in projection, or a bright part of the radio
halo neither behind or in front of the cluster (case 3). The
polarization properties and the trend of the polarization angle versus
$\lambda^2$ can help in distinguish among these three situations, and
we will discuss them in the following.
\begin{itemize}
\item{{\bf Case 1:}our galaxy acts like a Faraday screen, similarly
  to what we observe for the foreground source F, so we expect to
  obtain RM $\sim$ 10s rad/m$^2$.}
\item{{\bf Case 2:} the ICM acts like a Faraday screen and the polarization
  angle rotates following Eq. \ref{eq:rm}. }
\item{ {\bf Case 3:} in this case the situation is much more
  complicated. The trend of $\Psi$ versus $\lambda^2$ may result from
  complex geometries (see Burns 1966). In this case
      the rotation does not originate in an external Faraday screen,
    Eq.~\ref{eq:rm} does not hold anymore, and obtaining information
    about the magnetic field from Faraday rotation requires detailed
    knowledge of the ICM distribution and properties.} 
\end{itemize}

In Fig.~\ref{fig:RMrelic} we show fits of $\Psi$ versus $\lambda^2$
obtained in the relic-filament region. The poor agreement between the
data and the simple linear model suggests that the Faraday rotation is
not occurring in a Faraday screen, thus favouring scenario (3)
above. The observed trends of $\Psi$ versus $\lambda^2$ are also
incompatible with internal Faraday rotation generated by a uniform
slab (see Burn 1966).  
 Although strong beam depolarization could affect the trend of
  $\Psi$ versus $\lambda^2$, present data favour the third scenario
  and suggest that the relic-filament is actually a polarized filament
  belonging to the radio halo and not a radio relic connected to a
  peripheral merging shock.

\begin{figure}[!ht]
 \includegraphics[width=0.45\textwidth]{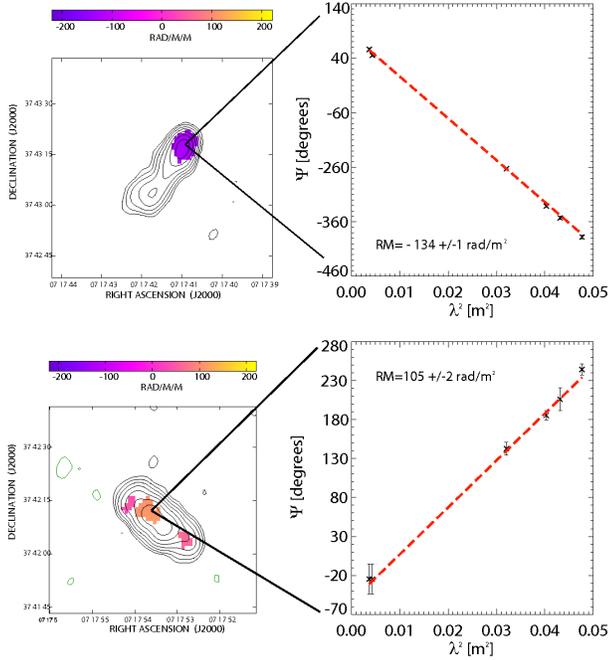}
   \caption{Left: Observed rotation measure images obtained for the source B
     (top panel) and C (bottom panel). Right panel: fit of the
     Polarization angles $\Psi$ versus $\lambda^2$, demonstrating that
     the Faraday rotation is occurring in a foreground screen. The
     fits are referred to a representative pixel in the two sources.}
\label{fig:RMsources}
\end{figure}
\subsubsection{\bf RM fit for the sources B and C}
In Fig.~\ref{fig:RMsources} we show the RM image and plots of $\Psi$
versus $\lambda^2$ obtained with PACERMAN for the two radio sources B
and C. The fits shown confirm that, as expected, the RM observed here
is due to an external Faraday screen. In Table \ref{tab:RMsource} we
report the observed RM for these sources ($RM_{obs}$) and the RM in
the cluster rest frame ($RM_{crf}$), that is given by
$RM_{crf}=RM_{obs}\times(1+z)^2$. Once the Galactic contribution is
subtracted, we obtain $\langle RM_{crf} \rangle$ values of --312$
\pm$1 rad/m$^2$ and 204$\pm$1 rad/m$^2$ for sources B and C,
respectively. As source C is associated with the BCG of a satellite
cluster embedded in the large-scale filament and visible in the X-ray
image, its RM is representative of the properties of that
cluster. There is no obvious concentration of X-ray emission around
source B, which is also located along the optical and X-ray
large-scale filament, but still within the virial radius of
MACS\,J0717 proper. A crude estimate of the gas density here is 7$ \pm
$1 10$^{-4}$ cm$^{-3}$.  Deriving the magnetic field from RM using
Eq.~\ref{eq:rm2} requires knowledge of the correlation-length scale
($\Lambda_{B}$) of the magnetic field (see Murgia et al.\ 2004). The
expectation value of the RM is in fact:
\begin{equation}
\langle RM^2\rangle \propto  \Lambda_B \int{(n_e(l)B_{//}(l))^2dl }
\label{eq:rm3}
\end{equation}
We have then to make some assumptions on $\Lambda_{B}$, and to fix the
limits of the integral in eq. \ref{eq:rm3}. If we assume that both gas
and magnetic field are uniform on a scale $\Lambda=$1 Mpc (i.e.\ the
projected distance from B to the center of the main cluster), and that
$\Lambda_B{=}1$ Mpc as well, we derive $\langle B\rangle
\sim$0.5$\mu$G. This value should be considered a lower limit to the
magnetic field strength. In fact, if we assume $\Lambda_{B}{=}200$ kpc
(the aproximate linear extent of source B), we obtain $\langle
B\rangle\sim$1.2 $\mu$G.  Although these estimates rely on several
assumptions, the values of both the RM and of the magnetic field
obtained are still higher than what has been observed previously in
sources located at such large distances from the cluster center (see
e.g.\ Clarke et al.\ 2004). Our findings thus indicate that the
magnetic field has already been amplified in these regions, possibly
by energetic phenomena associated with the complex merging history of
this cluster.

\begin{figure*}[ht]
\includegraphics[width=0.95\textwidth]{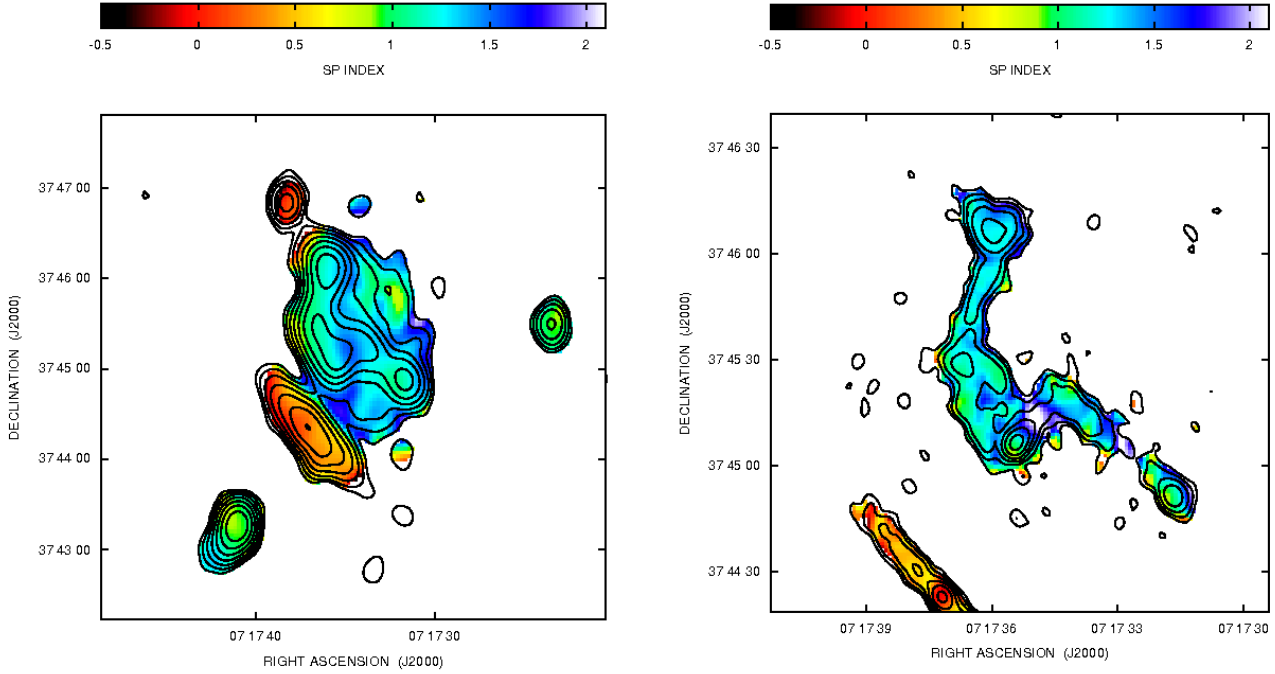}
 \caption{Left: spectral index image of the radio halo between 1.365
   GHz and 4.885 GHz. Contours refer to the radio emission at 4.860
   GHz. They start at 3$\sigma$ and are then spaced by a factor of
   2. The restoring beam FWHM is 18$''\times$14$''$. Right: spectral
   index image of the relic-filament between 1.365
   GHz and 4.885 GHz. Contours refer to the radio
   emission at 4.885 GHz. They start at 3$\sigma$ and are then spaced
   by a factor of 2. The restoring beam FWHM is 5$''\times$4$''$.}
\label{Fig:spix}
\end{figure*}

\section{Spectral index analysis}
\label{sec:spix}
\begin{figure}[h]
\includegraphics[width=0.49\textwidth]{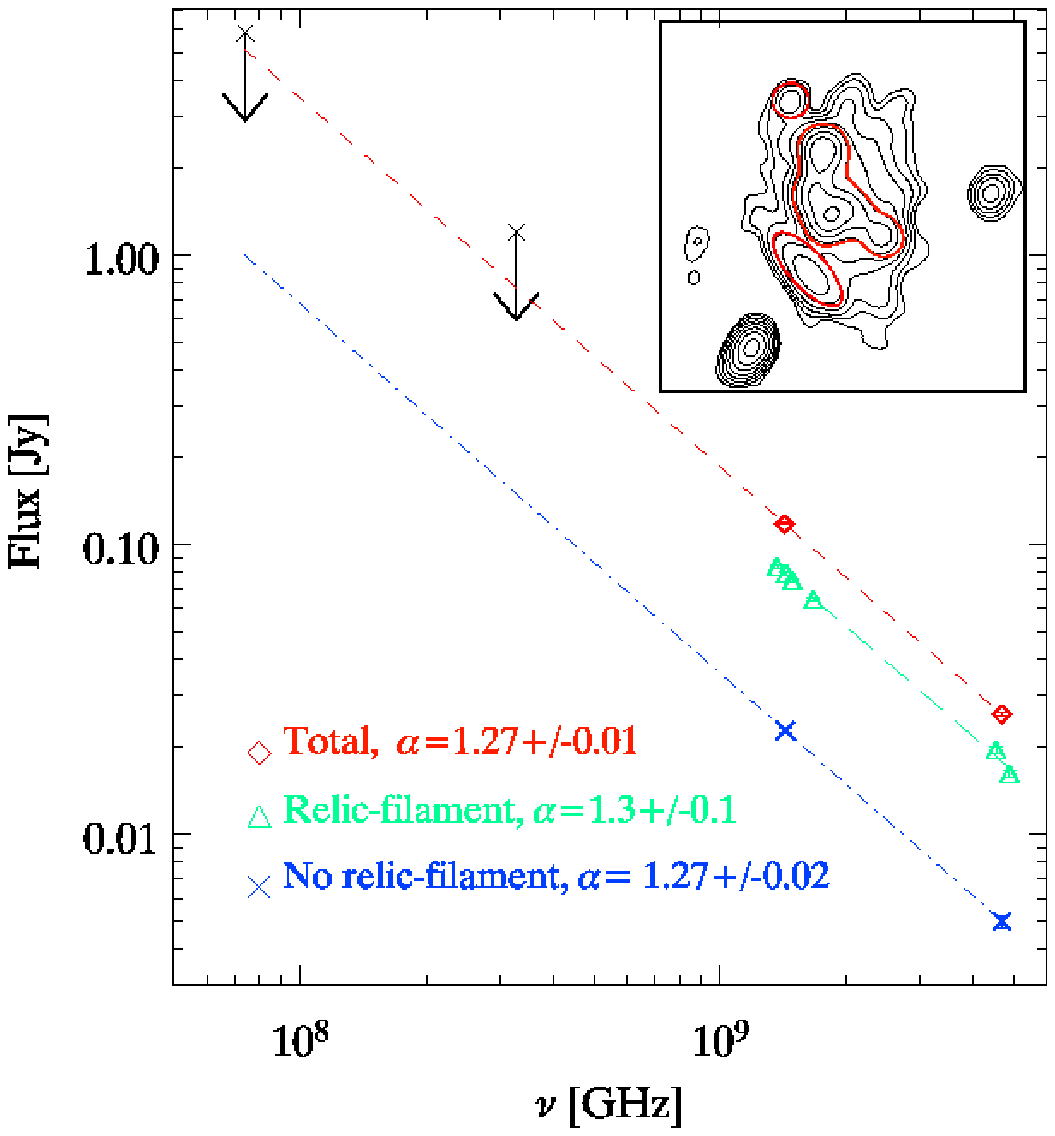}
\caption{Global spectral index fit for the relic-filament (triangles),
  for the halo once the relic-filament is subtracted (crosses) and for
  the total ICM emission (diamonds) between 20 cm and 6 cm. Arrows
  represent the upper-limits derived from the VLSS and WENSS. Bars
  represent 3$\sigma$ errors. In the inset red lines illustrate the
  regions that have been masked in order to obtain the spectral index
  of the halo once the relic-filament is subtracted.}
\label{fig:spixTot}
\end{figure}

\begin{figure}
\includegraphics[width=0.45\textwidth]{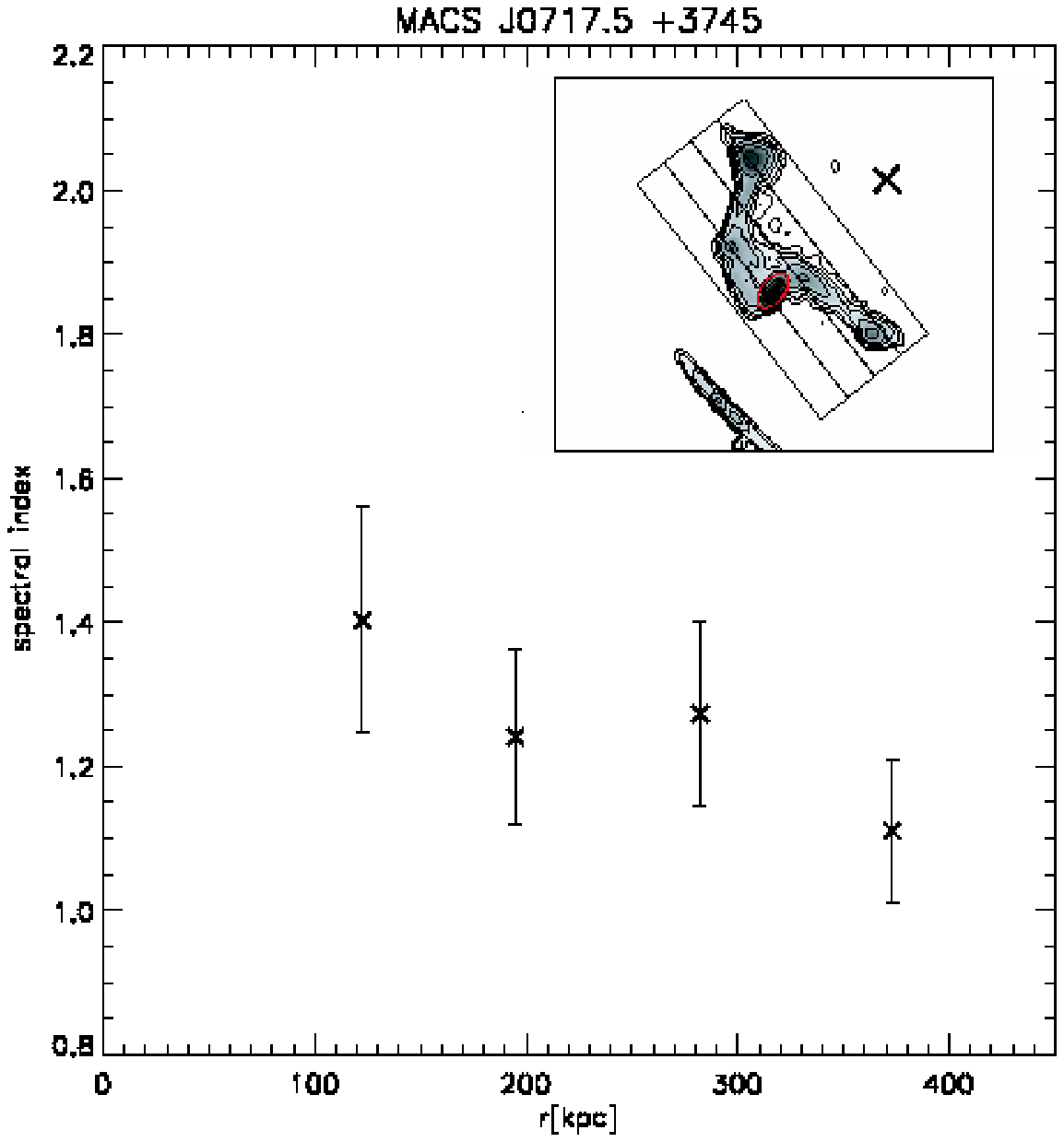}
 \caption{Spectral index profile between 1.365 and 4.885 GHz as a
   function of the distance from the cluster center. In the top right
   inset the displacement of the boxes is shown, the cross marks the
   location of the center. }
 \label{fig:Spix_prof}
\end{figure}

The analysis of the spectral index is useful to determine the
energetic spectrum of the emitting particles. In fact, it is well
known that if the emitting particle energy distribution follows a
power law $N(E)\propto E^{-\delta}$ the radio synchrotron spectrum
will result to be a power law as well $S(\nu)\propto\nu^{-\alpha}$,
with $\alpha=(\delta-1)/2$.\\ We derived the spectral index image by
comparing the high resolution images at 1.365 and 4.885 GHz and the
low resolution images at 1.425 and 4.860 GHz.  Spectral index images
are shown in Fig.  \ref{Fig:spix}. They were obtained considering only
pixels whose brightness is $>$3 $\sigma$ at both frequencies.  Images
at the same resolution were obtained using the same UV-range in order
to avoid any effect due to their different sensitivity to the large
angular structure.\\ We computed the global spectral index of the
relic-filament by fitting the integrated brightness as a function of
the frequency as:
\begin{equation}
\label{eq:spixFit}
LogS(\nu)=-\alpha Log(\nu).
\end{equation}
We obtained $\alpha\sim 1.3 \pm 0.1$ for the relic-filament, from high
resolution images, 1.27$\pm$0.02 for the radio halo once both the
relic-filament and the embedded sources are masked, and 1.27$\pm$0.01
for the entire diffuse radio emission (i. e. masking the embedded
sources only). The fits are shown in Fig. \ref{fig:spixTot}. We have
to consider here that the halo emission observed at 4.680 GHz is
likely affected by the Sunayev-Zeldovich effect, that causes a
decrement of the radio emission at 4.860 GHz and consequently a
steepening of the spectral index (see e. g.  Liang et al. 2000). \\ The
analysis performed in the previous Secs. indicates that the high
resolution emission previously classified as relic is likely a bright
region connected with the radio halo itself, rather than a peripheral
radio relic. Also the spectral index analysis agrees with this
hypothesis since the spectral index of the relic-filament alone, of
the halo once the relic-filament is masked and of the entire ICM radio
emission has the same value.\\ The analysis of the spectral index
profile in the relic-filament offers another possibility to check this
hypothesis.  The spectral index study of radio relics has been
performed so far on some clusters (see e. g. A2256, Clarke \& Ensslin
2006; A3667, Roettgering et al. 1997; A2744, Orr\'u et al. 2007; A521,
Giacintucci et al. 2008; A2345 and A1240, Bonafede et al. 2009), and
all of the present theoretical models require the presence of a shock
wave that either accelerates the particles from the thermal pool to
relativistic energies (Ensslin et al. 1998; Roettiger et al. 1999;
Hoeft \& Brueggen 2007) or compresses a region containing fossil radio
plasma amplifying the magnetic field and re-energizing the particles
so that they can emit radio wave again (Ensslin \& Gopal-Krishna
2001). In both of these cases a spectral steepening across the relic
main axis is expected if the relic is seen edge-on. The particles
accelerated or energized by the shock wave lose rapidly their energy
because of the combined effect of synchrotron and Inverse-Compton
losses. Their particle energy spectrum will thus steepen rapidly
giving rise to a radio spectrum that progressively steepens with the
distance from the current location of the shock.\\ In order to
investigate the presence of such a systematic trend, we integrated the
radio brightness at each frequency ($S_{\nu}$) in boxes of
$\sim$15$''$ in width. The associated error is then given by
$\sigma_{noise}\times \sqrt{N_{beams}}$, with $\sigma_{noise}$ being
the rms noise of the radio image, and N$_{beams}$ the number of beams
sampled in each box. The boxes are parallel to the relic main axis,
and are shown in the inset of Fig. \ref{fig:Spix_prof}. The spectral
index in each box was computed by fitting Eq. \ref{eq:spixFit}. \\ The
value of $\alpha$ goes from 1.4$\pm$0.2 in the inner box to
1.1$\pm$0.1 in the outer box, with values of 1.2$\pm$ 0.1 and
1.3$\pm$0.1 in the internal boxes. Its trend does not show a clear
progressive steepening as in the case of the other relics cited
above. Although we cannot exclude that this is due to ad-hoc
projection effects, this result agrees with the hypothesis that the
relic-filament is part of the more extended emission that is detected
with low resolution observations, i.e. it is a bright filament
belonging to the radio halo itself.\\ Thus, in the following analysis we
will consider the flux emitted by the whole extended structure (low
resolution emission +relic-filament), excluding only the contribution
of the embedded sources. We will refer to this whole emission as
  halo.\\

\begin{table}
\centering
  \begin{tabular}{cccc}
  \hline
Freq  & Beam &  Flux  & LLS  \\
 GHz  & $''\times''$& mJy & Mpc\\
\hline
1.425 &21$\times$18& 118$\pm$5 &    1.5\\
4.860 &18$\times$14& 26$\pm$1 & 1.0 \\
0.325 &84$\times$54&1.2$\pm$0.5$\times$10$^3$&1.8 \\
0.740&80$\times$80 & 5.8$\pm$0.6$\times$10$^3$& 1.3 \\
\hline
\end{tabular}
 \label{tab:Halo} 
  \caption{Radio Halo parameters: data at 74 and 325 MHz refer to the
    entire radio emission from the cluster, data at 1.425 and 4.860 
    GHz refer to the whole extended structure excluding only the contribution of the embedded
sources.}
\end{table}

\subsection{Spectral index of the radio halo}

\begin{figure}
\includegraphics[width=0.45\textwidth]{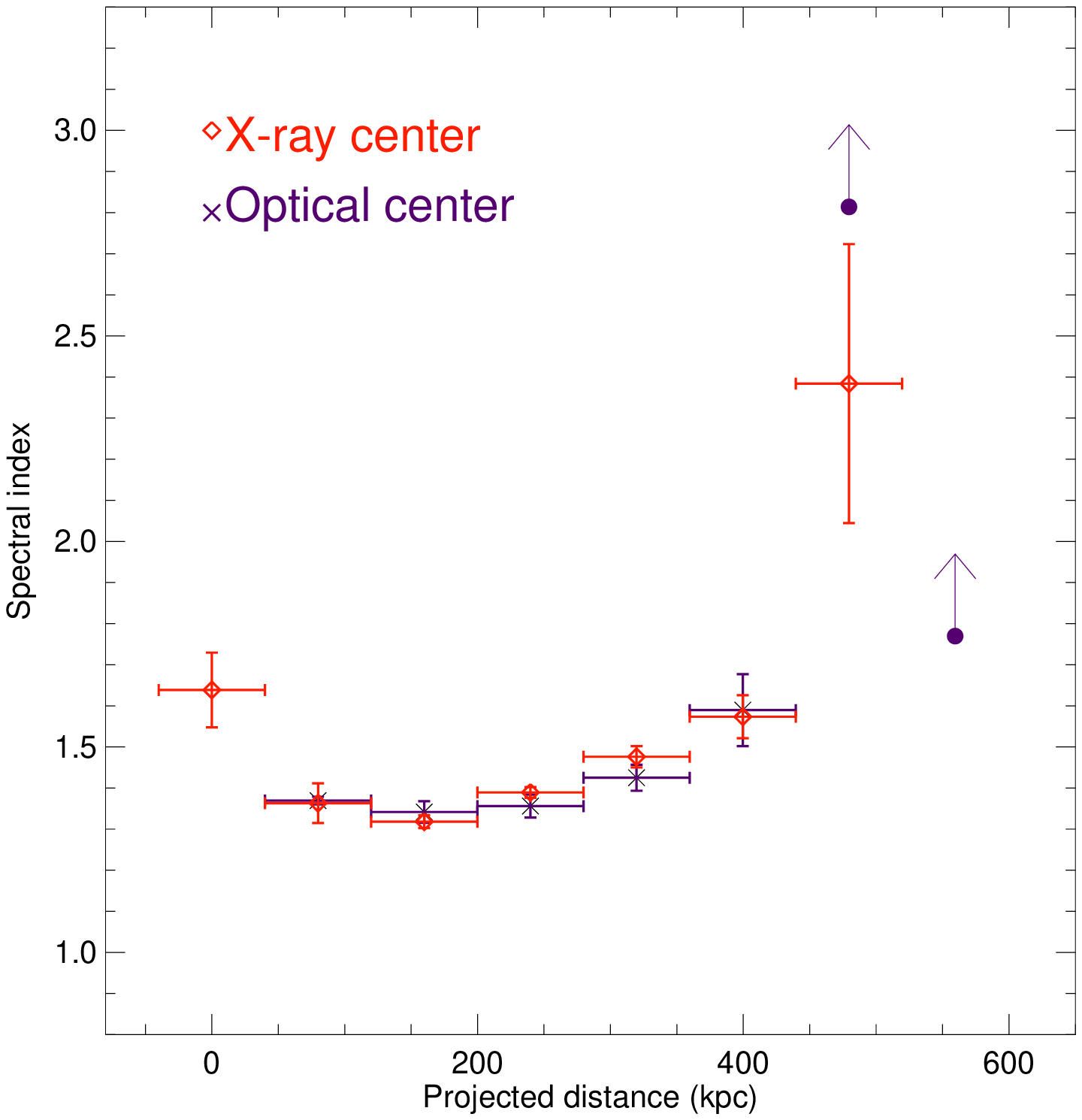}
 \caption{Spectral index profile of  the radio emission observed
     at low resolution (including the extended emission and the
     relic-filament region) computed in spherical shells of 10$''$ in
   width. Crosses represent the profile when shells are centered on
   the X-ray brightness peak, diamonds represent the profile once the
   shells are centered on the optical integrated light
   concentration. }
 \label{fig:plotSpixHalo}
\end{figure}

The total flux of the radio halo at 1.425 GHz and at 4.680 GHz are
$\sim$118 mJy and $\sim$26 mJy respectively. This makes MACS\,J0717
the most powerful radio halo ever observed.  Its radio luminosity at
1.425 GHz, once the k-correction is accounted for, is in fact
$\sim$1.6$\times$10$^{26}$ W$Hz^{-1}$.\\ The total flux have
been computed over the same area, excluding the regions where the
embedded sources are present, so that these values underestimate the
total flux of the radio halo and should be regarded as indicative
estimates. Due to the low resolution of both the VLSS and the WENSS
radio images, it is not possible to resolve the halo and the embedded
sources as different radio components, so that the radio flux measured
in those images can just be considered as upper limits to the halo
spectrum. In Fig. \ref{fig:spixTot} the flux density at the different
frequencies are shown. 74 MHz and 325 MHz limits are consistent with
the spectral index derived from the 1.425 and 4.680 GHz images.\\ The
distribution of the spectral index is an important observable in
a radio halo, since it depends on the shape of the electron spectrum
and on the magnetic field in which they emit. Regions of flatter
spectra indicate the presence of more energetic particles and/or
higher value of the magnetic field strength. A systematic variation of
the radio halo spectral index with radial distance from the cluster
center is predicted by re-acceleration models (see e. g. Brunetti et
al. 2001). In the re-acceleration scenario, in fact, particles are
accelerated up to a maximum energy that is given by the balance
between acceleration efficiency and energy losses. This results in a
break in the spectrum emitted by these electrons. The break frequency
depends on the acceleration efficiency and on the magnetic field
strength, so that allowing for a decrease of the magnetic field
strength with the radius, (see Sec. \ref{sec:MF}) a radial
steepening of the radio spectrum is expected, even in the case of a
radial constant acceleration efficiency (see Brunetti et al. 2001,
Brunetti 2003, Feretti et al. 2003 and references therein). This
radial steepening has been observed in some cases (see e. g.  Feretti
et al. 2004), while no steepening has been found in Abell 2744 (Orr\`u
et al. 2006). Here it is tricky to derive such a trend since the
cluster is in a very complex dynamical state. Ma et al. (2009) suggest
that the most massive structure is located at
RA$\sim$07$^h$17$^m$35$''$, DEC$\sim$37$^d$45$'$00$''$, that is not
coincident with the X-ray brightest region. The halo at 1.425 GHz is
more extended than at 4.680 GHz. In order to take this properly into
account in the spectral index analysis, we integrated the brightness
at 1.425 GHz and at 4.860 GHz in radial shells of $\sim$10$''$ in
width wherever the 1.4 GHz brightness is $>3\sigma$.  The associated
error is then $\sigma\times\sqrt{N_{beam}}$. In those shells where the
brightness is $>$3$\sigma$ in the 1.4 GHz image but $<$3$\sigma$ in
the 4.680 GHz image only lower limits on the mean spectral index can
be derived. We centered these shells on the X-ray cluster center and
on the optical condensation peak. The spectral index profile is shown
in Fig. \ref{fig:plotSpixHalo}. The flattest spectral index value is
in the shell that is 150 projected kpc from either the X-ray and the
optical center. Higher values of $\alpha$ are found in the shells with
radial distances $<$150 kpc and $>$200 kpc. A radial steepening is
thus detected centered on this point.

\begin{figure*}
\vspace{20pt}
\includegraphics[width=0.95\textwidth]{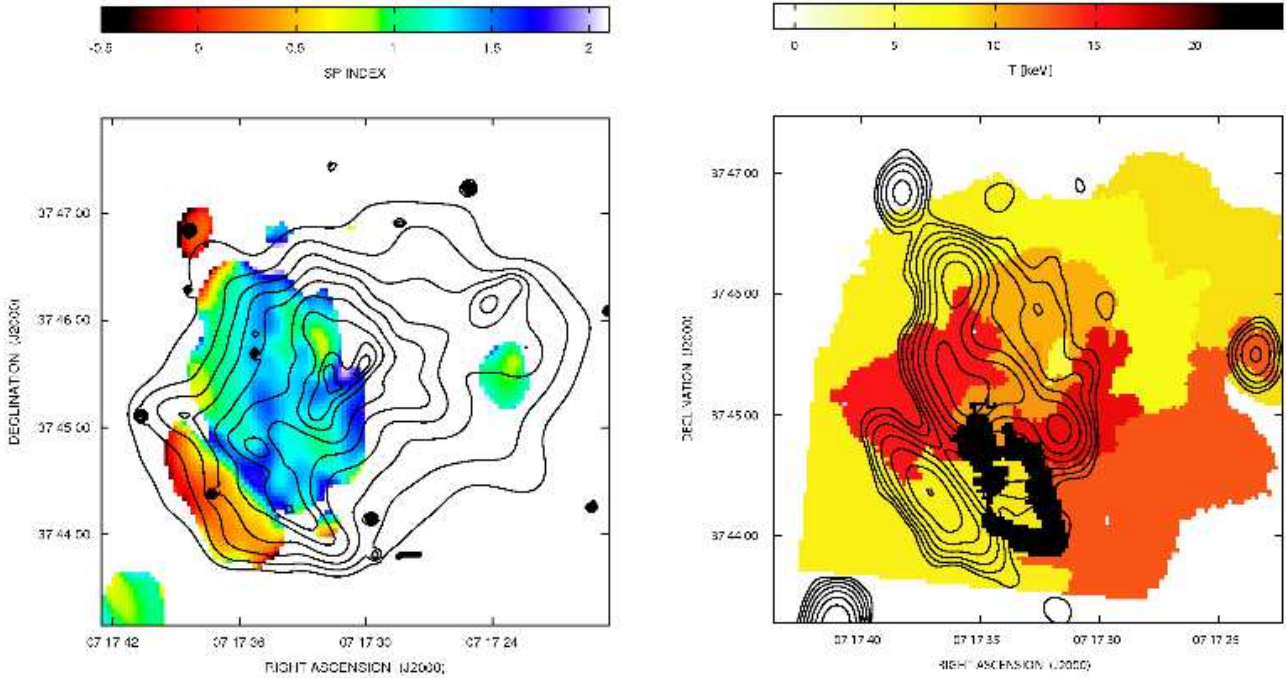}
 \caption{Left: Spectral index map is shown in colors, contours refer
   to the Chandra X-ray emission in the 0.7-5 keV band. Contours start at 0.06
   counts/s and are spaced by $\sqrt{2}$. Right: temperature map (from
   Ma et al. 2009) overlaid onto radio contours at 4.860 GHz. They
   start at 3$\sigma$ and are then spaced by a factor of 2. The
   restoring beam FWHM is 18$''\times$14$''$.}
 \label{Fig:halo_T}
\end{figure*}

\subsubsection{Spectral index - X-ray emission}
\begin{figure}
\includegraphics[width=0.45\textwidth]{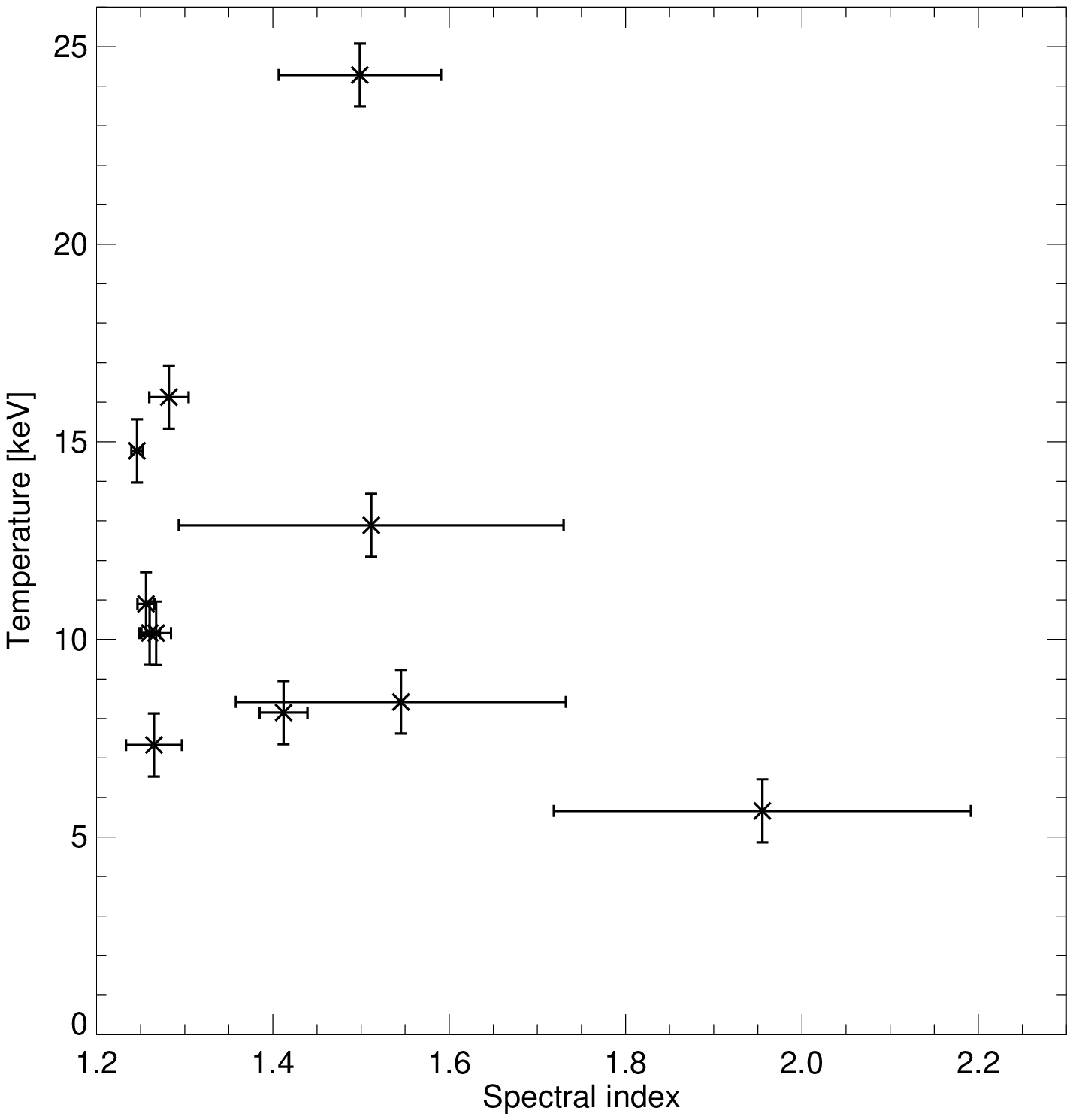}
 \caption{Temperature versus spectral index.}
 \label{Fig:alphaT}
\end{figure}

Radio properties of radio halos are linked to the properties of the
host cluster. This is directly demonstrated by several correlations
that have been found between the radio power and the cluster X-ray
luminosity (Bacchi et al. 2003), between the radio power and the
thermal gas temperature (Liang et al. 2000), between the radio power
and the total cluster mass (Govoni et al. 2001) and between the radio
spectral index and the thermal gas temperature (Feretti et al. 2004;
Orr\`u et al. 2006). Recently, Giovannini et al. (in prep) have
obtained a correlation between the radio halo integrated spectral
index and the average X-ray gas temperature for a large sample of
nearby radio halos. According to this correlation cold clusters (T $<$
8 KeV) show steep radio spectra (average spectral index = 1.7), while
hot clusters (T $>$ 10 KeV) show an average spectral index = 1.1 $-$
1.2. The radio spectrum of present radio halo with $\alpha\sim$ 1.27
confirms that flatter spectra are present in high temperature merging
clusters.  \\ A spatial comparison of the spectral index image and the
X-ray brightness image is shown in the left panel of
Fig. \ref{Fig:halo_T}. There is no clear correlation between the radio
halo spectral index and the X-ray emission. We note however that a
quite flat spectral feature is present in the NW region of the radio
halo, close to one of the X-ray peaks. With the help of the
temperature map (kindly supplied by C.J. Ma), we further investigate
the anti-correlation between the spectral index of the radio halo and
the ICM temperature. It is expected that flat spectrum regions have
higher temperature, since a fraction of the gravitational energy,
dissipated during mergers in heating thermal plasma, is converted into
re-acceleration of relativistic particles and amplification of the
magnetic field. In the right panel of Fig. \ref{Fig:halo_T} the
temperature map is shown (Ma et al. 2009).  A flatter region is
detected in the NW part of the radio halo, where the mean temperature
is 10.2$\pm$2.4 keV (Ma et al. 2009). However, in general, it is
difficult to match the patchiness morphology of the temperature map
with the spectral index map. \\ We also investigate the
anti-correlation between the ICM temperature and the radio spectral
index by computing the spectral index value in regions selected on the
basis of their temperature. We used the temperature map obtained after
the re-binning process (bottom panel of Fig. 2 in Ma et al, 2009). In
Fig. \ref{Fig:alphaT} the temperature versus the spectral index is
shown. We note that the coldest region is also characterized by the
most steep spectrum, but in general from this plot we can gather that
a correlation, if present, is weak in this cluster. We argue that it is
due to projection effects.

\section{ICM magnetic field}
\label{sec:MF} 
\begin{figure}
\includegraphics[width=0.45\textwidth]{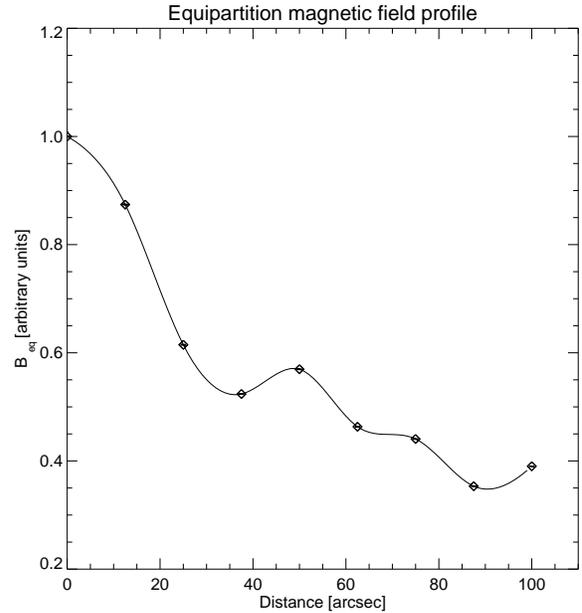}
 \caption{Equipartition magnetic field radial profile. It has been
   normalized to its central value. Regions where radio galaxies are
   present have been masked. }
\label{fig:haloBeqprof}
\end{figure}

The radio analysis performed so far can give important information
about the magnetic field in this cluster.\\ Under the assumption that
a radio source is in a minimum energy condition it is possible to
derive an average estimate of the magnetic field strength in the
emitting volume. We indicate with $\gamma$ the emitting particle
Lorentz factor. The synchrotron luminosity is calculated from a
low-energy cut-off of $\gamma_{min}=100$ and
$\gamma_{max}>>\gamma_{min}$ is assumed. We assume that the emitting
particle energy distribution is well represented by a power-law
$N(E)\propto E^{-\delta}$ with $\delta=2\alpha+1$ in this energy
range.  The volume of the halo is represented by an ellipsoid having
the major and minor axis equal to the maximum and minimum linear
extension of the relic, i. e. $\sim$1500 kpc and 1200 kpc
respectively, and the third axis is assumed to be the mean between
the major and the minor one. Under the assumption that magnetic field
and relativistic particles fill the whole volume of the source and
that the energy content in protons is the same as the one in electrons
we find that the equipartition magnetic field is $\sim$1.2 $\mu$G, in
agreement with other values found in the literature.\footnote{Standard
  equipartition estimate, i. e. obtained computing the radio
  synchrotron luminosity in a fixed range of frequency (10 MHz - 10
  GHz) would give $B_{eq}\sim$0.6$\mu$G. We report this value to
  compare this estimate with other given in the literature for other
  radio-sources. However it has been pointed out by Brunetti et
  al. (1997) that this approach is not self-consistent since electron
  energy corresponding to a fixed frequency depends on the magnetic
  field value.}\\ Under equipartition conditions, if we assume that
$\gamma_{min}$ is constant with cluster radius we obtain that
\begin{equation}
j_{\nu}\propto  B^{(\delta+5)/2}
\end{equation}
where $j_{\nu}$ is the synchrotron emissivity at the frequency $\nu$,
B is the magnetic field modulus and $\delta$ is the slope of the
particle energy distribution.\\ We used the deprojected brightness
profile of the radio halo at 1.425 GHz and computed the equipartition
magnetic field radial profile. This is shown in
Fig.\ref{fig:haloBeqprof}. The deprojected brightness profile was
computed assuming spherical symmetry with respect to the radio peak
brightness. The equipartition magnetic field decreases by a factor
$\sim$2.5 from the center to the periphery of the cluster. If the
cluster magnetic field decreases with radius as:
\begin{equation}
B(r)=B_0\left[1+\frac{r^2}{r_c^2}\right]^{\frac{-3\mu}{2}}
\end{equation}
it is possible to reproduce the equipartition magnetic field profile
assuming $\mu=1.1$, i.e. assuming that the magnetic field profile
scales as the gas density profile.  Once $\mu$ is fixed, it is
possible to derive the value of $B_0$ necessary to reproduce the
magnetic field equipartition estimate. We obtain that $B_0=$3 $\mu$G
averaged over the halo emitting volume ($\sim$ 1.1 Mpc$^3$) can
reproduce the equipartition magnetic field estimate.\\ The detection
of polarized emission reveals important information about the magnetic
field structure in this cluster.  Radio halos are intrinsically
polarized, since the synchrotron process generates linearly polarized
emission. However, in the ICM the emitting plasma is mixed with the
thermal one, so due to the Faraday Rotation significant depolarization
may occur. Moreover, radio halos have a low surface brightness, and
high resolution observations are often unable to detect them; if the
magnetic field is tangled on scales smaller than the beam size, the
observed emission will be further depolarized (beam
depolarization). These two effects can explain why polarized emission
from radio halos is usually non-detected. \\ The presence of polarized
emission here indicates that the magnetic field fluctuates on scales
as large as the beam, that is 130 kpc. Murgia et al. (2004) have
demonstrated that if the magnetic field power spectrum\footnote{The
  magnetic field power spectrum is modelled as
  $|B(\Lambda)|^2\propto\Lambda^{n}$, where $\Lambda$ is the
  fluctuation scale in the real space and n is the power spectrum
  spectral index} is steep enough (n$>$3) and the outer scale of the
magnetic field fluctuation is larger than few hundreds kpc, it is
possible to detect polarized emission from radio halos.\\ Radio halos
are expected to be generated by re-acceleration of particles by Fermi
II processes that occur in the ICM after merger events (see
e. g. Cassano \& Brunetti, 2005; Cassano, Brunetti \& Setti, 2006).
The energy is injected at large spatial scales and then a turbulent
cascade is generated. This cascade converts the energy into motions at
smaller and smaller scales until the dissipation scale is reached. The
analysis performed on this cluster suggests that it could be in a
young phase after the merger. The magnetic field is thus still ordered
on sufficient large scale and its polarized emission can be detected
even with low resolution observations.

\section{Conclusions}
\label{Sec:res}

The main result of this work is the discovery of a giant radio halo in
a massive galaxy cluster at z$\sim$0.55. The radio halo of MACS\,J0717
is the most distant radio halo ever observed and the most powerful
one. Moreover, it is only the second halo for which polarized emission
has been detected.  We here summarize the results of our analysis:
\begin{itemize}
\item{The cluster MACS\,J0717+3745 shows a complex radio morphology
    that reflects the complex dynamical state revealed by X-rays and
    optical studies.}
\item{A powerful radio halo has been observed, emission from which is
    detected here for the first time at 1.425 and 4.680 GHz. With a
    radio power of $\sim$1.6$\times$10$^{26}$ W$Hz^{-1}$ at 1.4 GHz, it is
    the most powerful radio halo ever observed.  Its spectral index is
    steep ($\alpha=$1.27$\pm0.01$), in agreement with results obtained
    for other radio halos found at lower redshifts. Our detection of
    diffuse radio emission from MACS\,J0717 at $z{=}0.55$ indicates
    that the ICM is already magnetized at this redshift.}
\item{We argue that the bright radio emission visible in
  high-resolution images and previously classified as relic is more
  likely a bright, polarized filament connected with the radio
  halo. This feature, in fat, lies at the center of the cluster and of
  the radio halo. The trend of the polarization angle $\Psi$ versus
  $\lambda^2$ indicates that the Faraday rotation originates in a
  region where a morphologically complex mix of thermal and
  non-thermal gas is present, resulting in a poor match both with the
  $\lambda^2$ law expected in the case of a Faraday screen and with
  the simplified model expected in the case of a uniform slab.
  Although the effects of depolarization cannot be taken into account
  trivially, we find the most plausible scenario to be one in which
  this filamentary emission is embedded in the central cluster
  region. The radio emission from this filamentarye structure is polarized at
  $\sim$8\% at 1.365 GHz and $\sim$17\% at 4.885 GHz, and the
  polarization image does not show any discontinuity between the radio
  halo and this relic-filament structure.  The spectral-index profile
  further reinforces our interpretation since no clear steepening is
  observed across the main axis, as would be expected if the emission
  were caused by a peripheral shock wave. We also note that the mean
  spectral index of the relic-filament and of the radio halo are fully
  compatible within the small errors.}
\item{Low-resolution polarization observations at 1.425 GHz have shown
    that the polarized emission is not confined to the bright
    relic-filament observed at high resolution but extends to the
    innermost regions of the radio halo and to some regions in the
    outskirts. Following Murgia et al.\ (2004), this indicates that
    the power spectrum of the magnetic field is steep in this cluster,
    with a spectral index n$>$3, and that it must fluctuate on scales
    as large as $\sim$130 kpc.}
\item{Under the equipartition assumption, we derived the
    magnetic-field profile of the radio halo and found it consistent
    with predictions based on the assumption that the magnetic-field
    profile scales as the gas-density profile. Once the radial decline
    is fixed, a central value of $\sim$3$ \mu$G can also account for
    the magnetic-field equipartition estimate. }
\end{itemize}

\bigskip

{\bf Acknowledgements} We thank C.J.\ Ma for providing the ICM
temperature map for MACS\,J0717, and A. Mantz for helpful comments.
 We thank the anonymous referee for useful comments. GG is a
postdoctoral researcher of the FWO-Vlaanderen (Belgium).   HE
  gratefully acknowledges financial support from SAO and STScI under
  grants GO3-4168X and GO-09722/GO-10420, respectively. NRAO is a
facility of the National Science Foundation, operated under
cooperative agreement by Associated Universities, Inc. This work was
partly supported by the Italian Space Agency (ASI), contract
I/088/06/0, by the Italian Ministry for University and Research (MIUR)
and by the Italian National Institute for Astrophysics (INAF). This
research has made use of the NASA/IPAC Extragalactic Data Base (NED)
which is operated by the JPL, California Institute of Technology,
under contract with the National Aeronautics and Space Administration.
We acknowledge the WENSS team (http://www.astron.nl/wow/testcode.php)

\end{document}